\documentclass[twocolumn,floatfix,prb,showpacs]{revtex4}%

\usepackage{epsfig}
\usepackage{dcolumn}
\usepackage{bm}
\usepackage{amsmath}
\usepackage{graphicx}

\begin{document}

\title{Time-domain measurements of quasiparticle tunneling rates in a single-Cooper-pair transistor}

\author{O. Naaman}

\author{J. Aumentado}
\email{jose.aumentado@boulder.nist.gov}
\affiliation{National Institute of Standards and Technology, 325 Broadway, Boulder CO., 80305}

\date{January 31, 2006}

\begin{abstract}
We have measured, in real time, individual quasiparticle tunneling in a single Cooper pair transistor, using rf reflectometry on the supercurrent branch. We have extracted the even-to-odd and odd-to-even transition rates directly by analyzing the distributions of dwell times in the even and odd states of the transistor. We discuss both the measurement and analysis techniques and report on the temperature and gate dependence of the quasiparticle tunneling rates.
\end{abstract}

\maketitle

Superconducting single-charge devices form an important subclass of quantum circuits, of which perhaps the best known is the Cooper-pair box (CPB) qubit.\cite{Nakamura99,Vion02,Wallraff04} In the CPB and several similar circuits, uncontrolled quasiparticle (QP) tunneling in the system can degrade their performance, causing the operating points of these devices to shift stochastically on time scales comparable to the measurement times. In CPB qubits, this can have the ultimate effect of reduced visibility of coherent oscillations, and may become an important decoherence mechanism as coherence times improve.\cite{Lutchyn05} In a related circuit, the single Cooper-pair transistor (SCPT), this uncontrolled `poisoning' of the supercurrent can be a limiting factor in its use as an electrometer, while QP tunneling in Cooper-pair pumps\cite{Aumentado03,Toppari04} severely inhibits their applicability as metrological current sources. While this problem has already been studied in some detail with other methods, in this paper we present rf measurements of SCPTs that allow us unprecedented access to the dynamics of QP tunneling in this class of circuits. As shown below, we have been able to monitor this process in the time domain and, by analyzing the resulting telegraph signal, determine the quasiparticle tunneling rates in a straightforward manner. 

Quasiparticle poisoning has been extensively investigated in the past both theoretically and experimentally in the SCPT.\cite{Averin92,Schon94,Tuominen92,Lafarge93,Joyez94,Amar94,Aumentado04,Mannik04} In those experiments the relative population of the odd (an excess QP on the island) and even (no excess QP on the island) states of the transistor was inferred, either by measuring the average charge on the island or by measuring the distribution of the current $I_{\text{sw}}$ at which the transistor switches out from its supercurrent branch. These experiments however \textit{do not} measure the even-odd and odd-even transition rates, $\gamma_\text{eo}$ and $\gamma_\text{oe}$ directly, whereas these rates are of particular interest in the context of charge qubit decoherence. We note that switching current measurements can in principle provide direct information about QP tunneling rates,\cite{Mannik04} but are sensitive only to QP tunneling events that \textit{reduce} the critical current of the transistor: at gate charges $n_g\equiv C_gV_g/e\in[0.5,1.5]$ these are even$\rightarrow\:$odd transitions ($C_g$ and $V_g$ are, respectively, the gate capacitance and voltage); for $n_g\in[-0.5,0.5]$ these are odd$\rightarrow\:$even transitions. Furthermore, inference of tunneling rates from these measurements is complicated by the evolution of the system during the current ramp.
\begin{figure}[b]
\epsfxsize=2.7in
\epsfbox{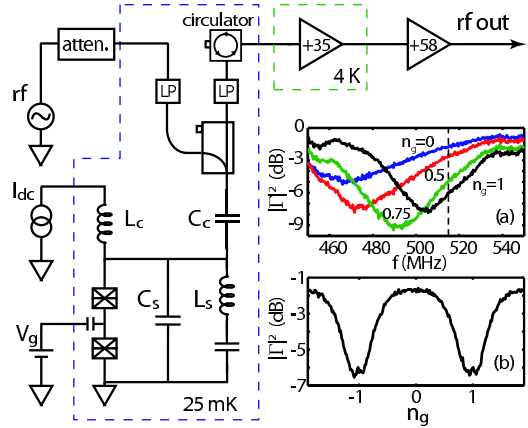}
\caption{\label{circuit} (Color online) A schematic of the experimental setup. RF power is fed to the circuit via a directional coupler after passing several stages of attenuation and low-pass (LP) filtering. The reflected power is amplified at 4~K, with additional amplifiers at room temperature. $C_c=3.3$ pF, $L_s=18$ nH, and $C_s=6.2$ pF. A 22 pF capacitor in series with $L_s$, and an rf choke $L_c=820$~nH allow dc biasing of the transistor. Inset (a): $|\Gamma|^2$ vs.\ frequency for device H, at different gate charges. Inset (b): $|\Gamma|^2$ vs.\ $n_g$ at 515 MHz. The power incident on the resonator was $P_\text{in}=10^{-15}$\:W (-120\:dBm).}
\end{figure}

Here we present the first direct measurement of QP tunneling rates in a zero-biased SCPT using time-domain rf reflectometry with microsecond temporal resolution. In contrast to previous experiments, our measurement is sensitive to both even$\rightarrow\:$odd \textit{and} odd$\rightarrow\:$even transitions. By measuring the impedance of the transistor biased on its supercurrent branch, we can read out the charge on its island without switching to the voltage state, thus avoiding additional pair breaking in the device.\cite{Moriond} Reading out the charge in this manner also avoids the added complication of re-initializing the system after each switching event.

To perform these experiments we embedded the SCPT in a resonant circuit, whose resonance frequency is determined by the parallel combination of $C_s$, $L_s$ (Fig.\ \ref{circuit}), and the source-drain impedance of the transistor. The impedance of the zero-biased SCPT includes the Josephson inductance,\cite{Sillanpaa04} $L_J(n_g)\sim h/2eI_c(n_g)$, where $I_c(n_g)$ is the charge dependent critical current. We found that this impedance additionally contains a real part, $R_J(n_g)$, likely due to fluctuations in the Josephson phase induced by the device's finite temperature and the local environmental impedance.\cite{Kautz90,Koval04,Coffey00} An rf carrier signal probes the circuit near its charge-dependent resonance frequency so that the power reflected from the resonator is a strong function of the island charge. A measurement of the reflection coefficient $|\Gamma|^2=P_\text{out}/P_\text{in}$ is therefore a measurement of the charge on the island, Fig.\ \ref{circuit}(a), and its temporal evolution measures the charge dynamics in the transistor. The circuit parameters were chosen to set its resonance near 500 MHz, the center of our circulator and amplifier band, and the probe power $P_\text{in}$ was set low enough to keep $L_J$ in the linear regime. Also note that the design of the circuit allows us also to characterize the device by use of conventional dc techniques.

Here we show data from two devices fabricated on the same chip,\cite{NoteRemote} using double-angle Al deposition. By oxygen-doping the Al in the first deposition step, we have produced films with different superconducting gaps, $\Delta_1=225\:\mu$eV (20 nm thick) and $\Delta_2=190\:\mu$eV (40 nm thick). In the first SCPT, which we will refer to as device H, the film with the higher gap $\Delta_1$ formed the island of the transistor, and the leads had the smaller gap $\Delta_2$. In the second SCPT, which we call device L, the gap of the island ($\Delta_2$) was lower than that in the leads ($\Delta_1$).\cite{Aumentado04} Otherwise the two devices were nearly identical: for device H, $E_c=170\:\mu$eV, $R_\text{N}=21.2\:\text{k}\Omega$, and for device L, $E_c=177\:\mu$eV, $R_\text{N}=21.3\:\text{k}\Omega$. Despite the similarity of the two devices, their different gap profiles lead to strikingly different QP `poisoning' dynamics~\textemdash~device L tends to trap quasiparticles on its island far more effectively than device H.\cite{Aumentado04}

The eigenenergies of the transistor's Hamiltonian, and correspondingly its critical current, are $2e$ periodic in the external charge applied to the island.\cite{JoyezThesis} Quasiparticle tunneling onto and off of the island shifts the net island charge by one electron, offsetting the modulation characteristic by half a period. Measurements of the transistor on time scales greater than the characteristic poisoning time will consequently show an average of its properties at $n_g$ and $n_g\pm1$, weighted respectively by the even and odd state probabilities. Faster measurements will show transitions between these charge parity states. 

In device H we observed no QP tunneling events and it appears `poisoning free'. However, from the device parameters and the QP trapping model of Ref.\ \onlinecite{Aumentado04}, we do expect fast QP transitions to occur near $n_g=1$ and conclude that their rates fall outside the bandwidth of our measurement. Our maximum measurement bandwidth\cite{bw_note} of 3~MHz  sets an upper bound on the lifetime of the odd parity state of the transistor, $\tau_\text{o}\leq0.3\;\mu$s. In addition, we observe no significant averaging of the reflected power away from its expected $2e$ modulation [Fig.\ \ref{circuit}(b)], suggesting that the transistor is predominantly in the even parity state. This requires that the even state lifetime be greater than $\tau_\text{o}$ by at least an order of magnitude, $\tau_\text{e}\geq3\;\mu$s. 

\begin{figure}[t]
\epsfxsize=2.9in
\epsfbox{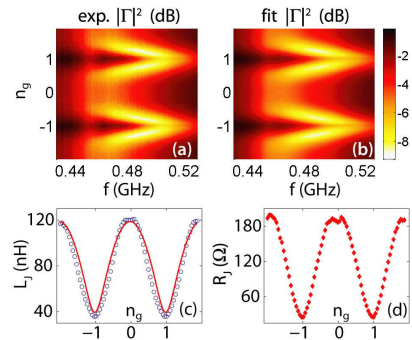}
\caption{\label{NWA} (Color online) (a) Intensity plot of the reflection coefficient $|\Gamma|^2$ (dB) in device H, measured as a function of gate charge and frequency at $T=25$ mK ($P_\text{in}=10^{-15}$\:W). (b) Calculated $|\Gamma|^2$ with $L_J(n_g)$ and $R_J(n_g)$ obtained from fits. (c) Fitted values of $L_J$ as a function of gate (circles), and the calculated $L_J(n_g)$ (line). (d) Fitted values of $R_J$.}
\end{figure}
We used the apparently `clean' $2e$-periodicity of device H to verify the operation of our rf circuit. Figure \ref{NWA}(a) shows the measured reflection coefficient $|\Gamma|^2$ as a function of frequency and gate charge in device H. At each gate charge, we fitted the experimental $|\Gamma(\omega)|^2$ with our circuit model, allowing $L_J$ and $R_J$ to vary while holding the other circuit elements fixed close to their nominal values. The fitted response is shown in Fig.\ \ref{NWA}(b). Figure \ref{NWA}(c) shows $L_J(n_g)$ obtained from the fits (circles), overlaid with the expected charge dependence of Josephson inductance in this device, calculated as in Ref.\ \onlinecite{Sillanpaa04} (line). We found that the charge-dependent loss, $R_J(n_g)$, modulates between approximately 20\textendash200~$\Omega$ [Fig.\ \ref{NWA}(d)], values that are somewhat lower than the phase diffusion resistance we measured at dc.\cite{note_diffusion} For the purpose of the present experiments, our simple model for the SCPT's impedance\,\textemdash~an inductor in series with a resistor\textemdash~agrees with the data sufficiently well; a full description of this impedance at arbitrary temperature, frequency, ac drive, and dc bias will require further experimental study.\cite{Coffey00} As an electrometer, the device's best charge sensitivity was $\delta q=5.2\times10^{-5}\;e/\sqrt{\text{Hz}}$, as estimated from the sideband signal-to-noise ratio with a small amplitude signal (0.025 $e$ rms) at $f=1.5$ MHz applied to its gate.\cite{Aassime01_2} These numbers are similar to the performance of conventional (dc-biased) rf-SET's,\cite{Thalakulam04} and are likely to improve with careful design of the rf circuit and the use of lower noise cryogenic amplifiers.

\begin{figure}[t]
\epsfxsize=2.7in
\epsfbox{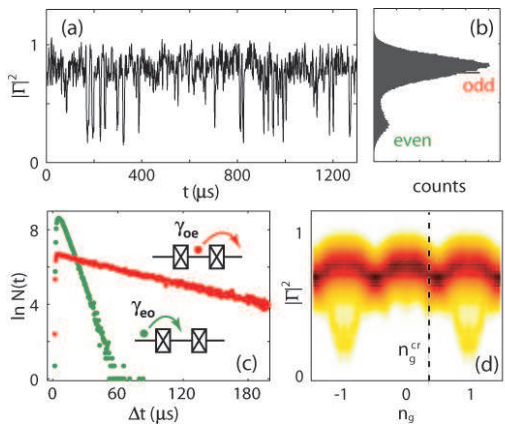}
\caption{\label{rt_combo} (Color online) (a) $|\Gamma|^2$ vs.\ time measured in device L at $n_g=1$. The resonator was probed at $f=505$ MHz with $P_\text{in}=1.6\times10^{-15}$\:W (-118\:dBm). (b) A histogram of the time record. The peaks correspond to the even and odd states of the SCPT. (c) The distribution of observed lifetimes in the even and odd states at $n_g=1$. (d) Intensity plot of time record histograms as in (b), as a function of $n_g$. Darker shades represents higher count number. Note the threshold for QP tunneling at $n_g^{cr}\sim0.35$ (dashed).}
\end{figure}

We now turn to device L, where the gap profile was reversed. Since the superconducting gap of the island in this device was smaller than that of the leads, the island trapped quasiparticles for longer periods of time than in device H, making the QP capture and emission process slow enough to be observed in the time domain. Figure \ref{rt_combo}(a) shows $|\Gamma|^2$ as a function of time, recorded at $n_g=1$. The reflected power switches randomly between two levels, seen clearly as distinct peaks in a histogram of the time record, Fig.\ \ref{rt_combo}(b), where the higher level corresponds to the odd state of the transistor with a QP trapped on the island, and the lower one corresponds to the even state. The typical time record in this experiment was one second long ($2\times10^6$ samples), encompassing well over $10^4$ tunneling events in each trace. In Fig.\ \ref{rt_combo}(d) we show the gate dependence of the time record histograms, revealing not only the rf response of the resonator [compare to Fig.\ \ref{circuit}(b)], but also the relative populations of the even and odd states as a function of $n_g$. Quasiparticle transitions were observed in a range of gate charges for which the even-odd energy difference $\delta E_\text{eo}=\delta E_\text{ch}+\delta\Delta>0$, where $\delta E_\text{ch}$ is the difference in charging energy and $\delta\Delta=35\;\mu$eV is the gap difference between the leads and the island.\cite{Aumentado04} The calculated threshold $n_g^{cr}\sim 0.35$, for which $\delta E_\text{eo}=0$, is shown in Fig.\ \ref{rt_combo}(d).

We determined the QP tunneling rates from the distribution of dwell times in each of the states, Fig.\ \ref{rt_combo}(c). These distributions are found by operating on the measured telegraph signal [Fig.\ \ref{rt_combo}(a)] with a change-point detection algorithm\cite{Lu03,Xiao04,Yuzhelevski00} and histogramming the time intervals between detected QP transitions. However, the experimentally observed transition rates, $\gamma^*_\text{oe}$ and $\gamma^*_\text{eo}$, which are obtained from exponential fits to the data in Fig.\ \ref{rt_combo}(c), underestimate the true rates in the underlying system, and must be corrected for the finite bandwidth of our measurement.\cite{Naaman06} QP tunneling events whose duration is shorter than the characteristic resolution time of our receiver cannot be detected, causing the lifetime histograms to be artificially skewed towards longer times (see also Ref.\ \onlinecite{Hawkes90}). We have shown previously that in time-domain measurements with a finite-bandwidth receiver the observed rates are given by\cite{Naaman06} 
\begin{equation}
\label{obs_rate}
\gamma^*_\text{eo\,(oe)}=\frac{1}{2}\left(\lambda-\sqrt{\lambda^2-4\,\gamma_\text{r}\gamma_\text{eo\,(oe)}}\right),
\end{equation}  
where $\lambda=\gamma_\text{eo}+\gamma_\text{oe}+\gamma_\text{r}$, the bandwidth of the receiver is $\gamma_\text{r}=\tau^{-1}_\text{r}$, and $\gamma_\text{eo}$ and $\gamma_\text{oe}$ are the true even-odd and odd-even rates in the underlying system. For the data shown here, $\tau_\text{r}=5\pm1\:\mu$s.

\begin{figure}[t]
\epsfxsize=2.7in
\epsfbox{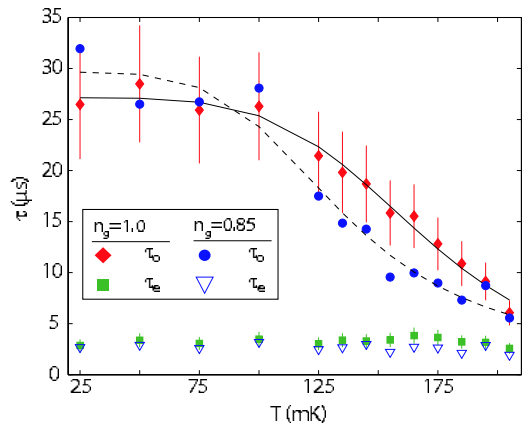}
\caption{\label{lifetime} (Color online) $\tau_\text{e}$ (squares, triangles) and $\tau_\text{o}$ (diamonds, circles) as a function of temperature, at $n_g=1$ and $n_g=0.85$. The data shown was corrected for finite bandwidth effects using the inverted Eq.\ (\ref{obs_rate}). The solid (dashed) line is a fit to the $n_g=1$ ($n_g=0.85$) data, incorporating the effects of electron heating by the probe rf signal, giving an activation barrier of $112\:\mu$eV ($48\:\mu$eV).}
\end{figure}

In Fig.\ \ref{lifetime} we show the temperature dependence of the even and odd state lifetimes of the transistor, $\tau_\text{e}=\gamma^{-1}_\text{eo}$ and $\tau_\text{o}=\gamma^{-1}_\text{oe}$, from which we can deduce the energetics of the poisoning process. The values shown in this figure were obtained from the observed rates by inverting Eq.\ (\ref{obs_rate}) to correct for finite bandwidth effects. The error bars are dominated by the uncertainty (20\:\%) in $\tau_\text{r}$. To verify the transformation in Eq.\ (\ref{obs_rate}) we varied the bandwidth of our measurement and extrapolated the observed lifetimes to $\tau_\text{r}\rightarrow0$ (infinite receiver bandwidth). Doing so, we obtained $\tau_\text{e}=3.0\pm0.5\;\mu$s and $\tau_\text{o}=26.0\pm0.7\;\mu$s at $n_g=1$ and $T=25$~mK, in agreement with the data in Fig.\ \ref{lifetime}.

From Fig.\ \ref{lifetime}, we see that $\tau_\text{e}$, the even-state lifetime, does not change significantly with temperature. This temperature independence implies that the even$\rightarrow$\:odd transition does not require activation, confirming the existence of a population of nonequilibrium quasiparticles on the leads.\cite{Aumentado04} In contrast, $\tau_\text{o}$ changes dramatically with temperature above $\sim80$~mK, and is clearly activated. At lower temperatures, where the electron-phonon thermal coupling is poor, the power dissipated in the device during the measurement causes the electron temperature and the odd-state lifetime to saturate. We fitted the odd state lifetime to $\tau_\text{o}\propto\exp(\Delta E/k_BT_e)$, where $T_e$ is the electron temperature given by $T_e^5=\left(P/\Sigma V+T^5\right)$.\cite{Wellstood94} Here $P$ is the power dissipated in the transistor, $\Sigma=2\times10^9$~W/m$^3$K$^5$, $V\sim3.2\times10^{-21}$ m$^3$ is the island volume, and $T$ is the temperature at the mixing chamber of our dilution refrigerator. The best fit at $n_g=1$ (solid line in Fig.\ \ref{lifetime}) is obtained with $P=1.6\times10^{-15}$ W, corresponding roughly to the incident rf power, and with an activation barrier $\Delta E=112~\mu$eV. For $n_g=0.85$, we find $\Delta E=48~\mu$eV, and the fit is obtained with a lower power $P=2.4\times10^{-16}$ W, consistent with a higher reflection coefficient of the circuit at this gate. 

If the odd$\rightarrow\:$even transition is due to the odd QP leaving the island, we would expect the activation barriers to be higher than our experimental values. We estimate that in our devices $\Delta E=\delta E_\text{eo}\simeq160\:\mu$eV and $140\:\mu$eV at $n_g=1$ and 0.85, respectively.\cite{2qp_note} These estimates, however, do not take into account quantum corrections to the even--odd energy difference due to virtual QP transitions\cite{Lehnert03,Thalakulam04} or higher-order (cotunneling) processes. For example, a second order process involving the simultaneous tunneling of a second QP into, and a Cooper pair out of the island may dominate the odd--even rates. We expect these corrections to be important in our devices, whose tunnel resistance $R_\text{N}<h/e^2$ puts them in the strong tunneling regime. We have also observed activation barriers that are smaller by nearly a factor of two than those expected from the orthodox theory, in switching current measurements in other SCPT devices having $R_\text{N}=18$~k$\Omega$ and $E_c\simeq115\:\mu$eV.

To conclude, we have measured single-Cooper-pair transistors using time-domain rf reflectometry on their Josephson branch. In one of the devices, whose island's superconducting gap was higher than that of its leads, we observed clean $2e$ periodicity of the reflected rf power as a function of gate charge. Quasiparticle transitions in this device, if they occurred, were too fast to be detected. In the other device, whose island had a smaller gap but was otherwise nearly identical to the first, we measured individual QP tunneling events in real time. We found the distribution of dwell times in the even and odd states of the transistor, and from it extracted the QP tunneling rates. This represents the first direct measurement of single-electron tunneling rates in a metallic transistor; these rates set a lower bound on the decoherence rates of superconducting charge qubits.\cite{Lutchyn05} The temperature independence of $\gamma_\text{eo}$ near $n_g=1$ indicates that the odd state of the transistor is energetically favorable. This observation, and the value of $n_g^{cr}$, the gate charge at which the odd state population becomes significant, are consistent with the picture of non-equilibrium QP population residing on the leads (Ref.\ \onlinecite{Aumentado04}). The temperature dependence of the odd$\rightarrow\:$even transition rates, however, yield activation barriers that are smaller than expected, possibly due to quantum corrections to the even--odd energy difference or to higher-order tunneling processes. 

We thank M.A.\ Sillanp\"{a}\"{a} and K.W.\ Lehnert for helpful comments.


\begin{thebibliography}{33}
\bibitem[{Not()}]{NoteRemote}
\bibinfo{note}{During measurement of one device, the other device was kept
  grounded to minimize `remote poisoning' effects, see Ref.\
  \onlinecite{Mannik04}}.
\bibitem[{bw_()}]{bw_note}
\bibinfo{note}{The demodulated signal was filtered to obtain a signal-to-noise
  ratio greater than 1. This reduces the effective bandwidth of the measurement
  below that of the resonator.}
\bibitem[{not()}]{note_diffusion}
\bibinfo{note}{We expect the effective loss to decrease with the number of
  phase slips per cycle of the ac drive}.
\bibitem[{2qp()}]{2qp_note}
\bibinfo{note}{The barriers for tunneling of a second QP onto the island are
  $\sim140\;\mu$eV (at $n_g=1$) and $195\;\mu$eV (at $n_g=0.85$).}
\end{thebibliography}

\end{document}